\documentclass[lettersize,journal]{IEEEtran}
\usepackage{amsmath,amsfonts}
\usepackage{algorithmic}
\usepackage{array}
\usepackage[caption=false,font=normalsize,labelfont=sf,textfont=sf]{subfig}
\usepackage{textcomp}
\usepackage{stfloats}
\usepackage{url}
\usepackage{verbatim}
\usepackage{graphicx}
\usepackage{float}
\usepackage{booktabs}
\usepackage{multirow}
\usepackage{cite}
\usepackage{svg} 
\usepackage{graphicx} 
\usepackage{hyperref}
\usepackage{multirow}
\begin{document}

\title{Super-temporal-resolution Photoacoustic Imaging with Dynamic Reconstruction through Implicit Neural Representation in Sparse-view}
\author{Youshen Xiao,
Yiling Shi,
Ruixi Sun,
Hongjiang Wei,
Fei Gao*,~\IEEEmembership{member~IEEE,}
Yuyao Zhang*,~\IEEEmembership{member~IEEE,}

\thanks{Youshen Xiao, Ruixi Sun and Yuyan Zhang are with the School of Information Science and Technology, ShanghaiTech University, Shanghai 201210, China.E-mail: \{xiaoysh2023, sunrx2023, zhangyy8\}@shanghaitech.edu.cn}

\thanks{Yiling Shi is with the School of Biomedical Engineering, ShanghaiTech University, Shanghai 201210, China. E-mail: shiyl2023@shanghaitech.edu.cn}

\thanks{Hongjiang Wei is with the School of Biomedical Engineering and Institute of Medical Robotics, Shanghai Jiao Tong University, Shanghai, China. E-mail: hongjiang.wei@sjtu.edu.cn}

\thanks{Fei Gao is with the School of Biomedical Engineering, Division of Life Sciences and Medicine, University of Science and Technology of China, Hefei, Anhui, 230026, China, with the School of Engineering Science, University of Science and Technology of China, Hefei, Anhui, 230026, China, and with the Suzhou Institute for Advanced Research, University of Science and Technology of China, Suzhou, Jiangsu, 215123, China. E-mail: fgao@ustc.edu.cn}

\thanks{Corresponding authors: Fei Gao and Yuyao Zhang.}

\thanks{The code for this work is availble at \url{https://github.com/zhaowumian7/INR_for_DynamicPACT}}
}

\markboth{Journal of \LaTeX\ Class Files,~Vol.~14, No.~8, August~2021}%
{Shell \MakeLowercase{\textit{et al.}}: A Sample Article Using IEEEtran.cls for IEEE Journals}

\maketitle
\begin{abstract}
Dynamic Photoacoustic Computed Tomography (PACT) is an important imaging technique for monitoring physiological processes, capable of providing high-contrast images of optical absorption at much greater depths than traditional optical imaging methods. However, practical instrumentation and geometric constraints limit the number of acoustic sensors available around the imaging target, leading to sparsity in sensor data. Traditional photoacoustic (PA) image reconstruction methods, when directly applied to sparse PA data, produce severe artifacts. Additionally, these traditional methods do not consider the inter-frame relationships in dynamic imaging. Temporal resolution is crucial for dynamic photoacoustic imaging, which is fundamentally limited by the low repetition rate (e.g., 20 Hz) and high cost of high-power laser technology. Recently, Implicit Neural Representation (INR) has emerged as a powerful deep learning tool for solving inverse problems with sparse data, by characterizing signal properties as continuous functions of their coordinates in an unsupervised manner. In this work, we propose an INR-based method to improve dynamic photoacoustic image reconstruction from sparse-views and enhance temporal resolution, using only spatiotemporal coordinates as input. Specifically, the proposed INR represents dynamic photoacoustic images as implicit functions and encodes them into a neural network. The weights of the network are learned solely from the acquired sparse sensor data, without the need for external training datasets or prior images. Benefiting from the strong implicit continuity regularization provided by INR, as well as explicit regularization for low-rank and sparsity, our proposed method outperforms traditional reconstruction methods under two different sparsity conditions, effectively suppressing artifacts and ensuring image quality.
\end{abstract}
\section{Introduction}
\IEEEPARstart{P}{hotoacoustic} imaging is a non-invasive biomedical imaging technique that integrates the rich contrast of optical imaging with the superior tissue penetration depth of ultrasound\cite{Zhou2016TutorialOP,Wang2008TutorialOP}. As a hybrid imaging technology, photoacoustic tomography (PACT) visualizes chromophores within biological tissues by converting absorbed light energy into acoustic energy\cite{Wang2009MultiscalePM}. Currently, PACT has three main implementation methods: Photoacoustic Microscopy (PAM), Photoacoustic Computed Tomography (PACT), and Photoacoustic Endoscopy (PAE). 
Among these, PACT is commonly used for dynamic imaging and serves as an imaging modality that provides tomographic images of biological tissues. 
In PACT, photon-induced acoustic waves, referred to as photoacoustic waves, are detected by an ultrasonic transducer array. The detected signals are used to reconstruct the initial pressure of the target tissue through inversion algorithms. Commonly used reconstruction algorithms include the Delay-And-Sum (DAS) method \cite{Matrone2015TheDM}, the Time Reversal (TR) method \cite{Treeby_2010}, and the Universal Back-Projection (UBP) method \cite{Xu2005UniversalBA}.

In medical and preclinical applications, the ability to monitor dynamic physiological processes is crucial for understanding disease progression and developing new therapies\cite{Kagadis2010InVS, Franc2008SmallAnimalSA}. Currently, this monitoring capability is largely dependent on a variety of medical imaging techniques. Ultrasound imaging allows real-time observation of blood flow and cardiac function\cite{Wang2022RealTimeUD, Kasai1987RealTimeTB}; the aim of dynamic magnetic resonance imaging (MRI) is to capture the dynamics of moving organs or fluctuations in blood oxygen levels, which necessitates a rapid imaging process at the expense of spatial resolution \cite{Fusco2021BloodOL, Tofts1999EstimatingKP}; dynamic computed tomography (CT) is a suitable tool for quantitatively monitoring the blood perfusion process in real-time, providing critical data on blood flow, volume, and tissue vascularization \cite{Madhavan2022LateralDD}; and positron emission tomography (PET) is used to detect changes in metabolic activity\cite{Sari2022FirstRO}. Compared to the imaging modalities mentioned above, PACT has emerged as a strong contender in dynamic imaging, offering high contrast and resolution, non-ionizing radiation, functional imaging capabilities, and real-time imaging. It is now commonly applied for real-time monitoring of blood flow and oxygenation levels \cite{am2023SpatiotemporalIR,Manohar2019CurrentAF,Upputuri2017DynamicIV,Lin2018SinglebreathholdPC}.


Although PACT technology has great potential in medical and biological imaging, current dynamic PACT techniques still face numerous limitations. Achieving high spatial and temporal resolution in two-dimensional or three-dimensional imaging requires a large number of acoustic sensors and high-repetition-rate laser sources. While these factors improve image quality, they also significantly increase system cost and hardware complexity. Although multiplexing PA signals can reduce the number of DAQ channels and system cost, this approach decreases imaging speed. To address this, sparse sensor arrays have become a common compromise in dynamic photoacoustic imaging. This method effectively reduces system cost and improves imaging speed by decreasing the number of sensors. However, the reduction in sensor count inevitably introduces severe image artifacts, affecting the accuracy of image reconstruction and the reliability of subsequent analysis. These artifacts not only degrade final imaging quality but also challenge the feasibility of PACT technology in practical applications. Therefore, achieving an optimal balance between system cost, temporal resolution, and image quality remains a critical issue in dynamic PACT technology.

In dynamic PACT research, traditional image reconstruction methods primarily employ reconstruction algorithms from static PACT, such as DAS and UBP, to independently reconstruct each data frame. This process is known as frame-by-frame image reconstruction (FBFIR). Although FBFIR methods are widely used in PACT research, they have two major limitations: first, they fail to fully exploit correlations between adjacent data frames, making them statistically and computationally suboptimal; second, these methods ignore temporal information in the measured data frames, limiting their ability to suppress measurement noise. Unlike FBFIR methods, spatiotemporal image reconstruction (STIR) techniques improve reconstruction accuracy by simultaneously estimating the entire image sequence. STIR has been widely applied in medical imaging fields such as CT, PET, and MRI, enabling high-precision reconstruction of dynamic objects even under sparse sampling conditions. Although STIR techniques have been applied in PACT, existing studies mostly focus on fully sampled tomographic data, aiming to improve reconstruction accuracy or reduce computational complexity. Recently, Cam et al. \cite{am2023SpatiotemporalIR} developed an STIR method for commercial volumetric PACT imaging systems, providing new insights into PACT applications.

In recent years, with the rapid development of deep learning in medical imaging, Seongwook et al. \cite{Choi2022DeepLE} proposed a deep-learning-based multi-parameter dynamic PACT method (DL-PACT). This method significantly enhances imaging quality and effectively reduces artifacts caused by low temporal resolution and sparse views. However, DL-PACT has certain limitations: first, its training process requires a large amount of data, making training costly and time-consuming; second, publicly available dynamic PACT datasets in the photoacoustic imaging field are scarce, further increasing the difficulty of applying supervised deep learning methods in PACT research.

Recently, a deep learning technique called Implicit Neural Representation (INR) has shown great potential in medical image reconstruction, registration, and analysis. Specifically, as a non-data-driven unsupervised approach, INR has demonstrated performance on certain reconstruction inverse problem tasks comparable to, or even exceeding, that of supervised deep learning methods \cite{Sitzmann2020ImplicitNR}. In dynamic imaging, Feng et al.\cite{Feng2022SpatiotemporalIN} proposed an INR-based method to improve dynamic MRI reconstruction from highly undersampled k-space data, demonstrating that it provides high-quality images with internal continuity. Zhang et al. \cite{Zhang2022DynamicCC} proposed a synchronous spatial-temporal INR method for dynamic cone-beam CT (CBCT) reconstruction. This approach models the unknown images and their motion to the evolution of a spatial and temporal multilayer perceptron (MLP), and iteratively optimizes the neuron weights of the MLP through the acquired projections to represent dynamic CBCT.

Inspired by recent advances in INR-based image reconstruction, this paper proposes
a novel unsupervised method to enhance the temporal resolution of dynamic PAI under sparse-view acquisition conditions. Our method treats the dynamic PA image sequence as a continuous function that maps spatiotemporal coordinates to corresponding image intensities. This function is parameterized by a MLP, which serves as an implicit continuity regularizer for dynamic PA image reconstruction. The MLP weights are learned directly from the spatial data consistency loss acquired by the ultrasound sensors, combined with explicit regularizers, without the need for a training database or any ground truth (GT) data. During inference, the reconstructed images can be simply queried from the optimized network with the same or denser spatiotemporal coordinates, allowing for sampling and interpolation of dynamic PACT at arbitrary frame rates; similarly, up-sampling to any resolution in the spatial domain is also possible. Experiments on the simulated Dynamic Abdominal PACT dataset of mouse demonstrate that the proposed method outperforms traditional reconstruction approaches under sparse-view conditions. The results show that even with only 64 ultrasound sensors, the peak signal-to-noise ratio (PSNR) can be improved by 18.20~dB to 18.76~dB. To further validate the effectiveness of the optimized representation function as an implicit regularizer for temporal continuity in dynamic photoacoustic image reconstruction, we conducted 4$\times$ temporal super-resolution tests using only one-quarter of the frames. The results confirm the effectiveness and generalizability of the proposed method.

The main contributions of this study are summarized as follows:

\begin{itemize}
\item This work introduces INR into dynamic photoacoustic imaging reconstruction for the first time, using it as an implicit temporal continuity regularizer to improve reconstruction quality under sparse-view conditions.

\item The proposed INR-based method is an unsupervised learning strategy that does not rely on external datasets or prior images for training. This makes the method highly generalizable to various ultrasound transducer array geometries (e.g., semi-circular, linear) and different imaging regions.

\item Without retraining the network, the proposed method achieves 4$\times$ temporal super-resolution using only one-quarter of the frames, demonstrating strong implicit temporal continuity and the potential for improving the temporal resolution of ultrafast PACT.
\end{itemize}
\section{Preliminary}
\subsection{PACT forward model}
In photoacoustic imaging, nanosecond laser pulses induce thermal expansion confined to the illuminated region, as thermal diffusion is negligible. Under the thermal and stress confinement conditions, the light source can be modeled as a Dirac delta function in time, and the resulting pressure wave in a homogeneous acoustic medium is governed by \cite{DenBen2012AccelerationOO}:

\begin{equation}
\frac{\partial^2 p(\mathbf{r}, t)}{\partial t^2} - c^2 \nabla^2 p(\mathbf{r}, t) = \Gamma H(\mathbf{r}) \frac{\partial \delta(t)}{\partial t},
\end{equation}

Here, \( c \) is the speed of sound in the medium, \( \Gamma \) denotes the Grüneisen parameter, and \( H(\mathbf{r}) = \mu_a(\mathbf{r}) U(\mathbf{r}) \) indicates the absorbed energy density, where \( \mu_a(\mathbf{r}) \) is the optical absorption coefficient and \( U(\mathbf{r}) \) is the local light fluence. Equivalently, (1) can be formulated as an initial value problem considering the homogeneous equation

\begin{equation}\begin{aligned}\frac{\partial^2p(\boldsymbol{r},t)}{\partial t^2}-c^2\nabla^2p(\boldsymbol{r},t)=0\end{aligned}\end{equation}

The pressure field \( p(\mathbf{r}, t) \) is governed by the following initial conditions:

\begin{align}
p(\mathbf{r}, t)\big|_{t=0} &= \Gamma H(\mathbf{r}),
\end{align}

\begin{equation}
\left. \frac{\partial p(\mathbf{r}, t)}{\partial t} \right|_{t=0} = 0,
\end{equation}

An analytical solution to the initial value problem defined in (2), (3) and (4) is provided by a Poisson-type integral of the form:

\begin{equation}p(\boldsymbol{r},t)=\frac{\Gamma}{4\pi c}\frac{\partial}{\partial t}\int_{S^{\prime}(t)}\frac{H(\boldsymbol{r}^{\prime})}{|\boldsymbol{r}-\boldsymbol{r}^{\prime}|}dS^{\prime}(t)\end{equation}

where \( S'(t) \) is a time-dependent spherical surface satisfying \( |\mathbf{r} - \mathbf{r}'| = ct \). The constant term outside the derivative in Eq.~(5) can be omitted if only the temporal profile of the pressure (in arbitrary units) is of interest. Furthermore, when the acoustic sources are confined to a plane, Eq.~(5) can be reduced to a 2D geometry, such that the integration is performed over a circular path \( L'(t) \), where \( |\mathbf{r} - \mathbf{r}'| = ct \). In this case, the pressure in arbitrary units is given by

\begin{equation}p(\boldsymbol{r},t)=\frac{\partial}{\partial t}\int_{L^{\prime}(t)}\frac{H(\boldsymbol{r}^{\prime})}{|\boldsymbol{r}-\boldsymbol{r}^{\prime}|}dL^{\prime}(t).\end{equation}

\subsection{Problem formulation}
The dynamic PACT acquisition process can be represented by the following linear model:

\begin{equation}
\mathbf{y} = \mathbf{Ax},
\end{equation}

Where $ \mathbf{A} \in \mathbb{R}^{(S \times F) \times (N \times N)} $ represents the forward model of PACT acquisition, as described in the previous section. $ \mathbf{x} \in \mathbb{R}^{(N \times N) \times T} $ denotes the discretized image matrix, and $ \mathbf{y} \in \mathbb{R}^{((S \times F) \times T) \times T} $ represents the measurement data. Here, $ S $ refers to the number of sensors, $ F $ is the number of sampling points, $ N $ indicates the size of the reconstructed image, and $ T $ represents the number of time frames during the acquisition process. The inverse problem involves solving for the unknown dynamic PACT image $ \mathbf{x} $ from the sensor data $ \mathbf{y}  $.

In sparse-view dynamic PACT acquisition, the measurement data $ \mathbf{y}  $ is undersampled to reduce the data volume and improve the system's temporal resolution, resulting in the dimensionality of $ \mathbf{y}  $ being much smaller than that of the image matrix $ \mathbf{x} $ (i.e., $ (S \times F) \ll (N \times N) $). Consequently, the matrix $ \mathbf{A} $ becomes rank-deficient, making dynamic PACT reconstruction an ill-posed problem. A typical solution is to formulate this problem as a regularized inversion problem, which can be expressed as follows:

\begin{equation}
    \arg \min_{\mathbf{x}} \| \mathbf{y} - \mathbf{Ax} \|_2^2 + \mathcal{R}(\mathbf{x}),
\end{equation}

where $\|\mathbf{y} - \mathbf{A}\mathbf{x}\|_2$ represents the data-fidelity term, ensuring that $\mathbf{x}$ remains consistent with $\mathbf{y}$. $\mathcal{R}(\mathbf{x})$ is the prior regularizer, helping target $\mathbf{x}$ reach optimal results under ill-posed conditions.

In other dynamic medical imaging modalities, such as MRI and CT, studies have shown that using sparsity and low-rank regularization as prior knowledge in both compressive sensing (CS)-based methods \cite{Lingala2011AcceleratedDM,zhao2012image,Otazo2015LowrankPS} and DL-based methods \cite{Huang2020DeepLP} can achieve state-of-the-art results in dynamic reconstruction. One example can be formulated as:

\begin{equation}
    \arg \min_{\mathbf{x}} \| \mathbf{y} - \mathbf{Ax} \|_2^2 + \lambda_D \| TV_t(\mathbf{x}) \|_1 + \lambda_L \| \mathbf{x} \|_*,
\end{equation}

where: $ TV_t(\cdot) $ denotes the temporal total variation (TV) operator as the sparsity regularization term; $\|\mathbf{x}\|_*$ represents the nuclear norm of the image matrix $\mathbf{x}$ (the sum of singular values), serving as the low-rank regularization term; $\lambda_D$ and $\lambda_L$ are the hyperparameters for low-rank and sparsity regularization, respectively. Previous works \cite{Lingala2011AcceleratedDM,zhao2012image} have demonstrated that, in the absence of GT, the objective function in (9) can be iteratively optimized to achieve excellent dynamic performance.
\section{Method}

\subsection{INR in dynamic PACT}
As an emerging unsupervised paradigm, INR have demonstrated significant potential in photoacoustic image reconstruction tasks \cite{shen2022nerp,zha2022naf,lin2023learning}. Within the INR framework, reconstruction can be achieved by applying a learnable continuous mapping function between spatiotemporal coordinates and the intensity of the image to be reconstructed:

\begin{equation}
    f : \mathbf{p} = (x, y, t) \in \mathbb{R}^3 \rightarrow I(\mathbf{p}) \in \mathbb{R},
\end{equation}

Here, $ I(\mathbf{p}) $ represents the intensity of the PACT image $ \mathbf{x} $ at the spatiotemporal coordinate $ \mathbf{p} $. The function $ f $ is parameterized as a MLP network $ \mathcal{M}_\theta $, where $ \theta $ denotes the learnable parameters of the network. Here, $ (x, y) $ represents the two-dimensional spatial coordinates ($1 \leq x, y \leq N$), and $ t $ represents the temporal coordinate ($1 \leq t \leq T$). 

\begin{equation}
    I(\mathbf{p}) = \mathcal{M}_\theta(\gamma(\mathbf{p})), 
\end{equation}

where the function $\gamma(\cdot)$ represents an encoding module, designed to capture specific features such as high-frequency components through positional encoding \cite{mildenhall2021nerf}, Fourier encoding \cite{tancik2020fourier}, or hash encoding \cite{Mller2022InstantNG} for efficient rendering. This process leverages the inherent capability of neural networks to learn along the manifold of images \cite{Rahaman2018OnTS}, effectively facilitating the reconstruction.

Consequently, the image $ \mathbf{x} $ is represented as $ \mathbf{x}_\theta \in \mathbb{R}^{(N \times N) \times T} $, which is organized into a Casorati matrix by inputting all fixed spatiotemporal Cartesian coordinates of the dynamic image into $ f_\theta $:

\begin{equation}
  \mathbf{x}_\theta=\begin{bmatrix}\mathcal{M}_\theta(\gamma(1,1,1))&\cdots&\mathcal{M}_\theta(\gamma(1,1,T))\\\vdots&&\vdots\\\mathcal{M}_\theta(\gamma(N,1,1))&\ddots&\mathcal{M}_\theta(\gamma(N,1,T))\\\vdots&&\vdots\\\mathcal{M}_\theta(\gamma(N,N,1))&\cdots&\mathcal{M}_\theta(\gamma(N,N,T))\end{bmatrix},
\end{equation}

Based on that, Equation (7) can be formulated as a fitting problem aimed at finding the optimal parameters $ \theta $ of the continuous mapping function $ f_\theta $.

\begin{equation}
    \arg \min_{\mathbf{\theta}} \| \mathbf{y} - \mathbf{Ax_\theta} \|_2^2 + \lambda_D \| TV_t(\mathbf{x_\theta}) \|_1 + \lambda_L \| \mathbf{x_\theta} \|_*,
\end{equation}

Here, (13) incorporates implicit continuity over the desired image sequence, along with explicit sparsity and low-rank regularization terms. The corresponding pipeline is illustrated in Fig. 1.

\begin{figure*}[t!]
\centering
\includegraphics[width=7.4in]{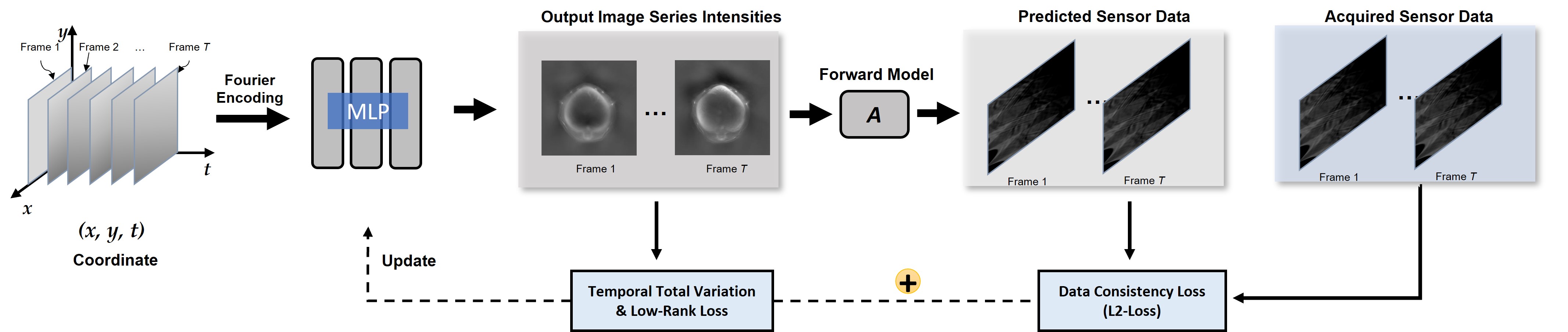}%
\label{pipelineaa}
\centering
\caption{Overview of the proposed method is as follows: All spatiotemporal coordinates are first input into a Fourier encoding and then fed into a MLP to output the signal intensity at each point as the initial pressure of the image. The predicted sparse sensor data is then obtained from the reconstructed image using a forward model. The difference between the predicted sparse sensor data and the actual acquired sparse sensor data is computed as the data consistency loss. In the loss function, two regularization terms are applied to the output image sequence: Total Variation in time and Low-rankness. The parameters in the hash grid and MLP are iteratively updated by minimizing the loss function.}
\end{figure*}

\subsection{Continuous mapping function with MLP and fourier encoding}

To learn the INR of dynamic PACT images, we employ a MLP \( \mathcal{M}_\theta \) that approximates the target reconstruction by fitting the sparsely sampled sensor measurements \( \mathbf{y} \). To improve the MLP's ability to represent high-frequency components of dynamic images, we adopt Fourier feature mapping \cite{tancik2020fourier}, which projects the spatiotemporal coordinates \( \mathbf{p} \in \mathbb{R}^d \) into a higher-dimensional space before passing them to the network. Formally, given an input coordinate \( \mathbf{p} \), the Fourier mapping is defined as:
\[
\gamma(\mathbf{p}) = \left[ \cos(2\pi \mathbf{B} \mathbf{p}), \sin(2\pi \mathbf{B} \mathbf{p}) \right]^T \in \mathbb{R}^{2L},
\]
where \( \mathbf{B} \in \mathbb{R}^{L \times d} \) is a matrix whose elements are sampled from a Gaussian distribution \( \mathcal{N}(0, \sigma^2) \), and \( L \) controls the number of frequency bands. This mapping allows the network to better capture fine-grained variations in both spatial and temporal dimensions. Compared with traditional MLP directly trained on raw coordinates, Fourier features introduce periodic basis functions that help mitigate the spectral bias of neural networks toward low-frequency functions. Consequently, the network gains enhanced capacity to represent the high-frequency structures commonly found in dynamic photoacoustic images. Moreover, Fourier mapping is a lightweight and general encoding scheme, requiring no additional learned parameters and providing consistent performance across different levels of sparsity.

\subsection{Loss functions}
We employ INR to handle dynamic PACT images, representing the dynamic image as a continuous implicit function as shown in Eq. 9. Subsequently, we formulate the problem of dynamic image reconstruction using the INR method as finding the optimal set of parameters \(\theta\) to minimizing the loss function:

\begin{equation}
  \mathcal{L}_{total}=\underbrace{\|\mathbf{y} - \mathbf{Ax_\theta} \|_{2}^{2}}_{\mathcal{L}_{DC}}+\lambda_{D}\underbrace{\|TV_{t}(\mathbf{x}_{\theta})\|_{1}}_{\mathcal{L}_{TV}}+\lambda_{L}\underbrace{\|\mathbf{x}_{\theta}\|_{*}}_{\mathcal{L}_{LR}} .
\end{equation}

Here, \( L_{\text{DC}} \), \( L_{\text{TV}} \), and \( L_{\text{LR}} \) represent the data consistency (DC) loss in the \((k, t)\) space, the temporal TV loss, and the low-rank (LR) loss, respectively.

We adopt the \( L_2 \) loss function as the DC term, to account for the fact that the amplitudes of low-frequency components are several orders of magnitude higher than those of high-frequency components. The temporal total variation operator \( TV_t(\cdot) \) is employed as a sparsity regularization term, while the nuclear norm of the image matrix \( \|\mathbf{x}_\theta\|_* \) is used as the low-rank regularization term. The contributions of these two regularization terms are controlled by hyperparameters \( \lambda_D \) and \( \lambda_L \), respectively. Prior studies \cite{zhao2012image} on dynamic image reconstruction in other modalities have demonstrated that this regularization strategy performs well even in the absence of GT.

\subsection{Implementation details}
The whole pipeline is illustrated in Fig.1. We use a small MLP with 3 hidden layers, each consisting of 256 neurons, followed by a ReLU activation function. The MLP outputs a single channel representing the initial pressure values of the PA image. The final layer uses a Sigmoid activation function instead of a ReLU activation function. For Fourier feature mapping, we choose \( L = 256 \) to ensure a sufficient level of frequency resolution for accurately capturing high-frequency details in the dynamic photoacoustic images.

During the optimization process, all spatiotemporal coordinates are grouped into a single batch with a batch size of 1. All coordinates are isotropically normalized to the range [0, 1] to facilitate faster convergence. The number of optimization iterations is set to 100-2000. We use the Adam optimizer\cite{Kingma2014AdamAM}. The learning rate starts at \(10^{-3}\) and is gradually decreased in each training phase until it reaches \(10^{-6}\).

Once the optimization process is complete, the continuous function \( f_{\theta} \) is considered a good representation of the underlying image sequence. The same batch of coordinates or a denser batch of coordinates is then fed into the INR network to output the image sequence. The proposed method is implemented in PyTorch and runs on a workstation equipped with an NVIDIA GPU (Titan RTX). All hyperparameters are chosen empirically and then fine-tuned based on experimental results.

\section{Experiments}

\subsection{Datasets}
\textit{1) The Dynamic Abdominal PACT dataset of mouse:} To evaluate the proposed method, we performed simulation experiments using a publicly available dynamic PACT dataset of a mouse \cite{liang2025organ}. The original images were acquired with a commercial small-animal multispectral PACT system (MSOT inVision128, iThera Medical, Germany), featuring a laser pulse duration of approximately 5\,ns and a repetition rate of 10\,Hz, enabling a minimum temporal resolution of 0.1\,s. The laser wavelength is tunable within the range of 680\,nm to 980\,nm. In this study, we selected 20 consecutive frames and cropped each frame to a spatial resolution of \(128 \times 128\). To simulate sensor acquisition under sparse-view settings, we generated sinograms using a forward model with two different levels of sparsity: 128 sensors and 64 sensors configurations.

\textit{2) In vivo Dynamic Finger Data:} To evaluate the applicability of the proposed method in real-world scenarios, we collected dynamic photoacoustic data from the finger of a healthy volunteer. The experiment was conducted underwater using a 128-channel PACT system to acquire multiple frames of raw PA signals, followed by offline processing. A pulsed laser with a wavelength of 720\,nm and a repetition rate of 20\,Hz was used for illumination. The laser light was delivered through a fiber bundle and uniformly distributed in a circular pattern around the transducer to ensure homogeneous illumination. The system's sampling rate was 40\,MSa/s, and the image reconstruction area was set to \(20\,\mathrm{mm} \times 20\,\mathrm{mm}\). Among all the collected data, 20 consecutive frames were selected for experimental validation. The raw data were then uniformly downsampled to simulate a sparse-view configuration with 64 sensors. All reconstructed images had a resolution of \(128 \times 128\).

\textit{3) In vivo Fish data:} To further evaluate the applicability of the proposed method in real-world scenarios, we conducted \textit{in vivo} tomographic photoacoustic imaging experiments on live fish. A dedicated whole-body small animal imaging scanner based on cross-sectional tomography geometry was used for data acquisition. PA signals were detected by a ring-shaped transducer array composed of 128 individual elements (center frequency: 2.5\,MHz, Doppler Ltd.) arranged in a circle with a radius of 30\,mm. A PACT system was employed to record the PA signals for image reconstruction. Illumination was provided by a pulsed laser with a wavelength of 720\,nm and a repetition rate of 100\,Hz. The laser light was guided by a fiber bundle and evenly distributed in a circular pattern around the transducer for uniform illumination. The data acquisition system operated at a sampling rate of 40\,MSa/s, and the reconstruction area was set to \(20\,\mathrm{mm} \times 20\,\mathrm{mm}\). Live fish were placed in a water tank, and as they swam freely, the scanner continuously captured their cross-sectional images at different positions. From the acquired data, a sequence of 100 consecutive frames was selected for experimental validation. The raw data were then uniformly downsampled to simulate sparse-view acquisition using 64 sensors.

\subsection{Compared methods}
In this study, we selected DAS \cite{Matrone2015TheDM}, UBP \cite{Xu2005UniversalBA} as baseline methods for detailed performance comparison. DAS is widely regarded as one of the most classical and commonly used beamforming algorithms in photoacoustic imaging. It reconstructs images by applying time-delay corrections to the received signals followed by summation. Due to its simplicity and high computational efficiency, it has been extensively applied in various PA imaging scenarios. UBP is a classical model-driven method that reconstructs images using an integral-based backprojection approach. One of its key features is the incorporation of a solid angle weighting factor to compensate for spatial response inconsistencies caused by non-uniform sensor distribution or limited view coverage. This improves the reconstruction accuracy and uniformity to some extent. For all baseline methods, their fundamental configurations (e.g., network architectures) were set according to the recommendations in their original publications to ensure fair comparisons and faithful performance reproduction.


To quantitatively assess performance, we use the structural similarity index (SSIM)\cite{Wang2004ImageQA} and PSNR to evaluate the performance of different methods. 

PSNR is defined as:
\begin{equation}
  PSNR=10\times log_{10}(\frac1{\|y-\hat{y}\|_2^2}),
\end{equation}

SSIM is defined as:
\begin{equation}
  SSIM= \frac{(2\mu_{y}\mu_{\hat{y}}+c_{1})(2\sigma_{y\hat{y}}+c_{2})}{(\mu_{y}{}^{2}+\mu_{\hat{y}}{}^{2}+c_{1})(\sigma_{y}{}^{2}+\sigma_{\hat{y}}{}^{2}+c_{2})} ,
\end{equation}

Where \( y \) and \( \hat{y} \) represent the GT and reconstructed images, respectively. \(\mu_y\) and \(\mu_{\hat{y}}\) are the mean intensities of \( y \) and \( \hat{y} \), \(\sigma_y\) and \(\sigma_{\hat{y}}\) are the variances of \( y \) and \( \hat{y} \), and \(\sigma_{y\hat{y}}\) is the covariance of \( y \) and \( \hat{y} \). The constants \( c_1 \) and \( c_2 \) were set to \(0.01^{2}\) and  \(0.03^{2}\), respectively. Both \( y \) and \( \hat{y} \) were normalized to the range [0, 1] based on the maximum and minimum values in the image sequence.

\begin{figure*}[t!]
  \centering
  \includegraphics[width=\textwidth]{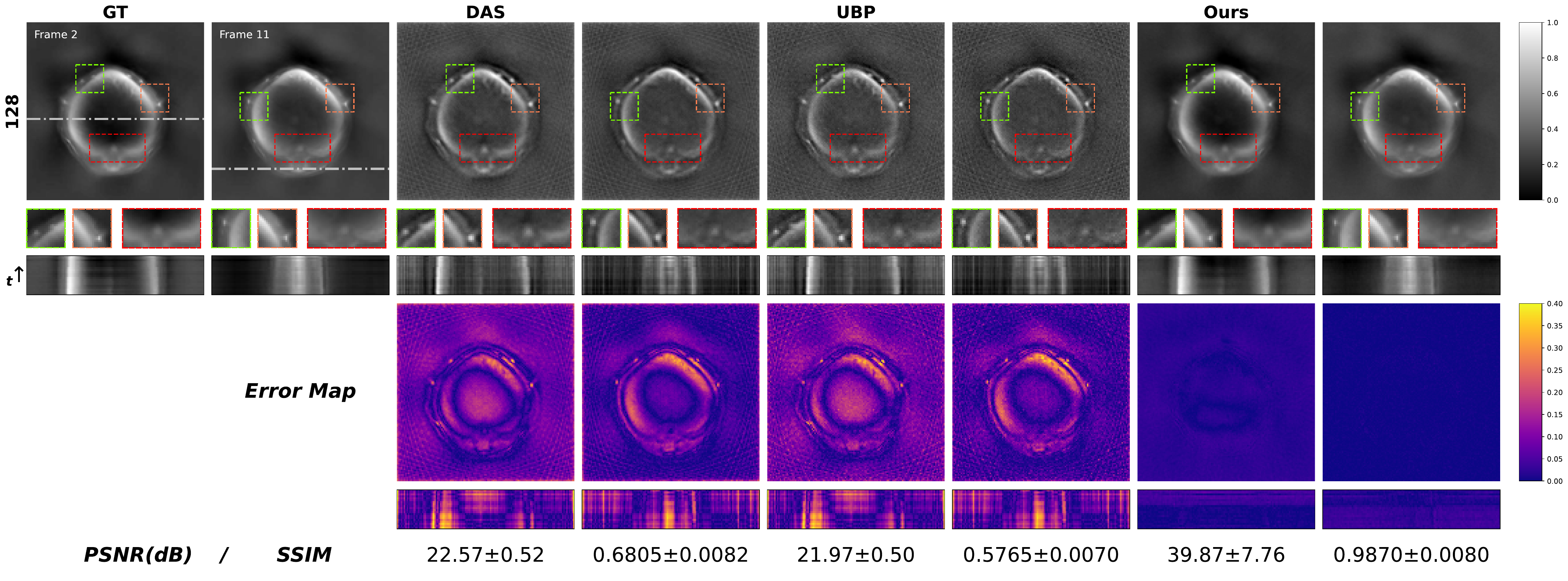}%
  \label{fig:example}
  \centering
  \caption{The reconstruction results on the simulated dynamic abdominal PACT data of mouse for DAS, UBP, and the proposed method (from left to right) are shown with 128 sensors. The enlarged view of the abdominal region is highlighted by yellow boxes, and the corresponding x-t image (extracted along the gray dashed line in the image) is also displayed to illustrate temporal consistency. It is noted that the proposed method provides superior reconstruction performance in terms of structural details and image fidelity. Error maps and average PSNR/SSIM metrics are displayed at the bottom. }
\end{figure*}
\begin{figure*}[t!]
  \centering
  \includegraphics[width=\textwidth]{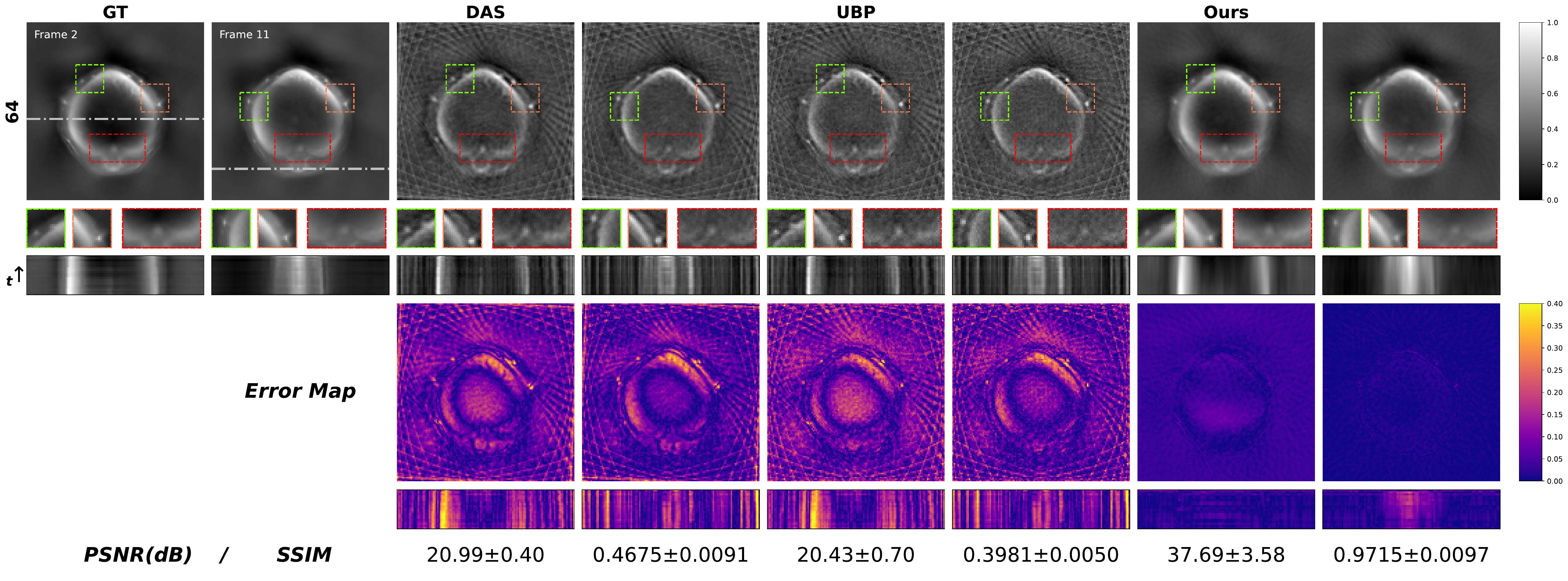}%
  \label{fig:example}
  \centering
  \caption{The reconstruction results on the simulated dynamic abdominal PACT data of mouse for DAS, UBP, and the proposed method (from left to right) are shown with 64 sensors. The enlarged view of the abdominal region is highlighted by yellow boxes, and the corresponding x-t image (extracted along the gray dashed line in the image) is also displayed to illustrate temporal consistency. It is noted that the proposed method provides superior reconstruction performance in terms of structural details and image fidelity. Error maps and average PSNR/SSIM metrics are displayed at the bottom. }
\end{figure*}

\subsection{Reconstruction performance comparisons}
\textit{1) The Dynamic Abdominal PACT Dataset of Mouse:}
Fig.2 and Fig.3 compare the reconstruction performance of different methods on the dynamic abdominal PACT mouse dataset with sparse views of 128 and 64 sensors. Visually, the images reconstructed by our proposed method exhibit better anatomical details and effectively remove artifacts. With 128 sensors, DAS and UBP methods can roughly reconstruct the abdominal shape, but they show blurring in fine details and contours (as indicated by the red, green, and orange boxes in the figure). Additionally, both DAS and UBP methods generate noticeable artifacts in the background region. As the sparsity increases, the reconstruction quality of DAS and UBP deteriorates further, with increased blurring and significantly more artifacts in the background. In contrast, our proposed method reconstructs images that closely resemble the true anatomical details, with clearer geometric edges and a clean background without artifacts. Our method also provides the highest temporal fidelity in the inter-frame dynamic images. The error maps between the reconstructed results and the GT further support our observations. Quantitatively, our proposed method performs the best. The average PSNR values on this dataset are: 39.87 dB with 128 sensors and 37.69 dB with 64 sensors. The average SSIM values are: 0.9870 with 128 sensors and 0.9715 with 64 sensors. These experimental results demonstrate the effectiveness of our proposed method in sparse reconstruction of dynamic simulation data.
\begin{figure*}[t!]
  \centering
  \includegraphics[width=\textwidth]{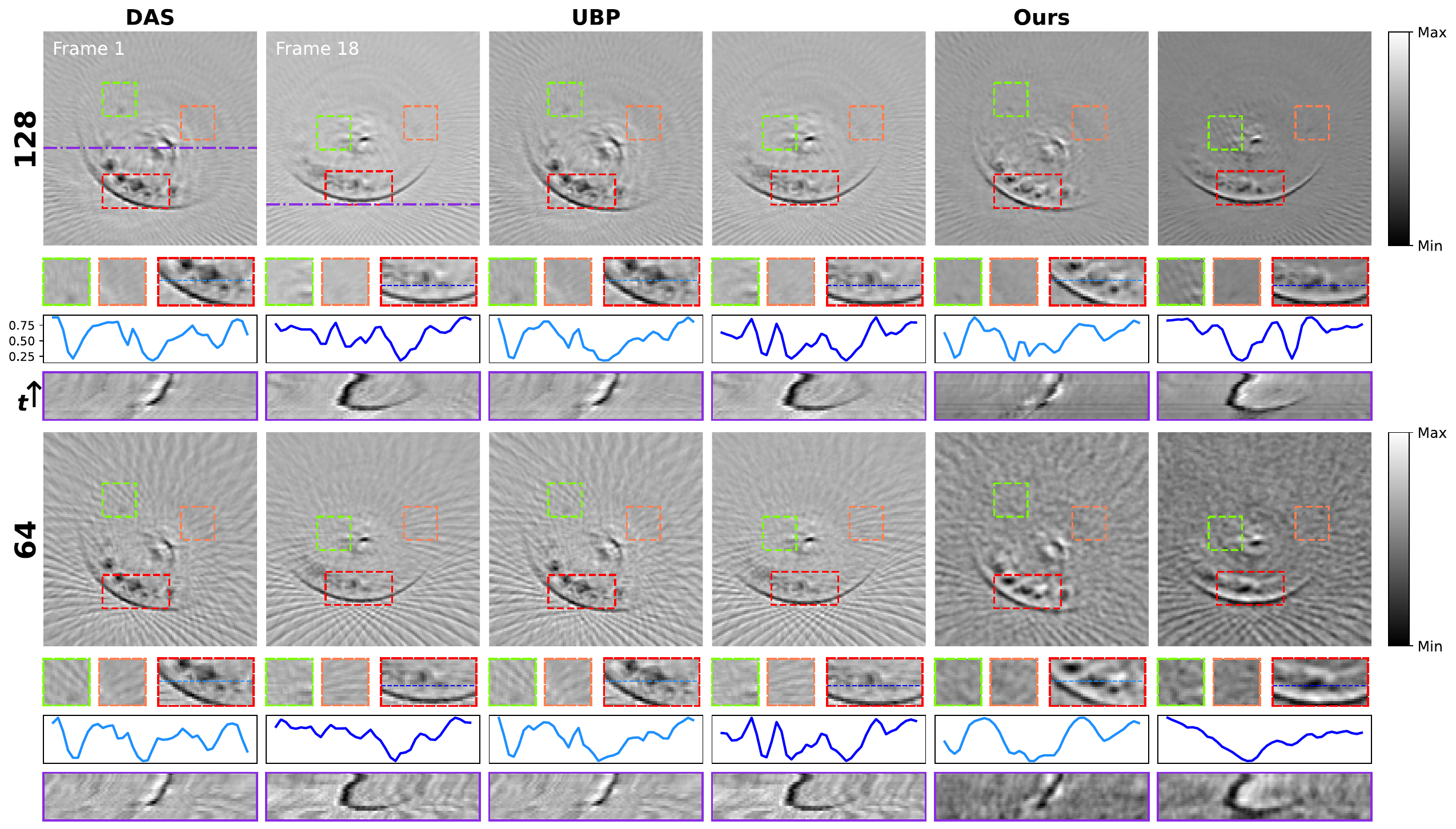}%
  \centering
  \caption{The reconstruction results on \textit{in vivo} finger data for DAS, UBP, TR, and the proposed method (from left to right) are shown with 16, 32, 64, and 128 sensors(projections). The enlarged view of the heart region is highlighted by yellow boxes, and it is noted that the proposed method provides superior reconstruction performance in terms of structural details. Error maps and average PSNR/SSIM metrics are displayed at the bottom.
  }
\end{figure*} 

\begin{figure*}[t!]
  \centering
  \includegraphics[width=\textwidth]{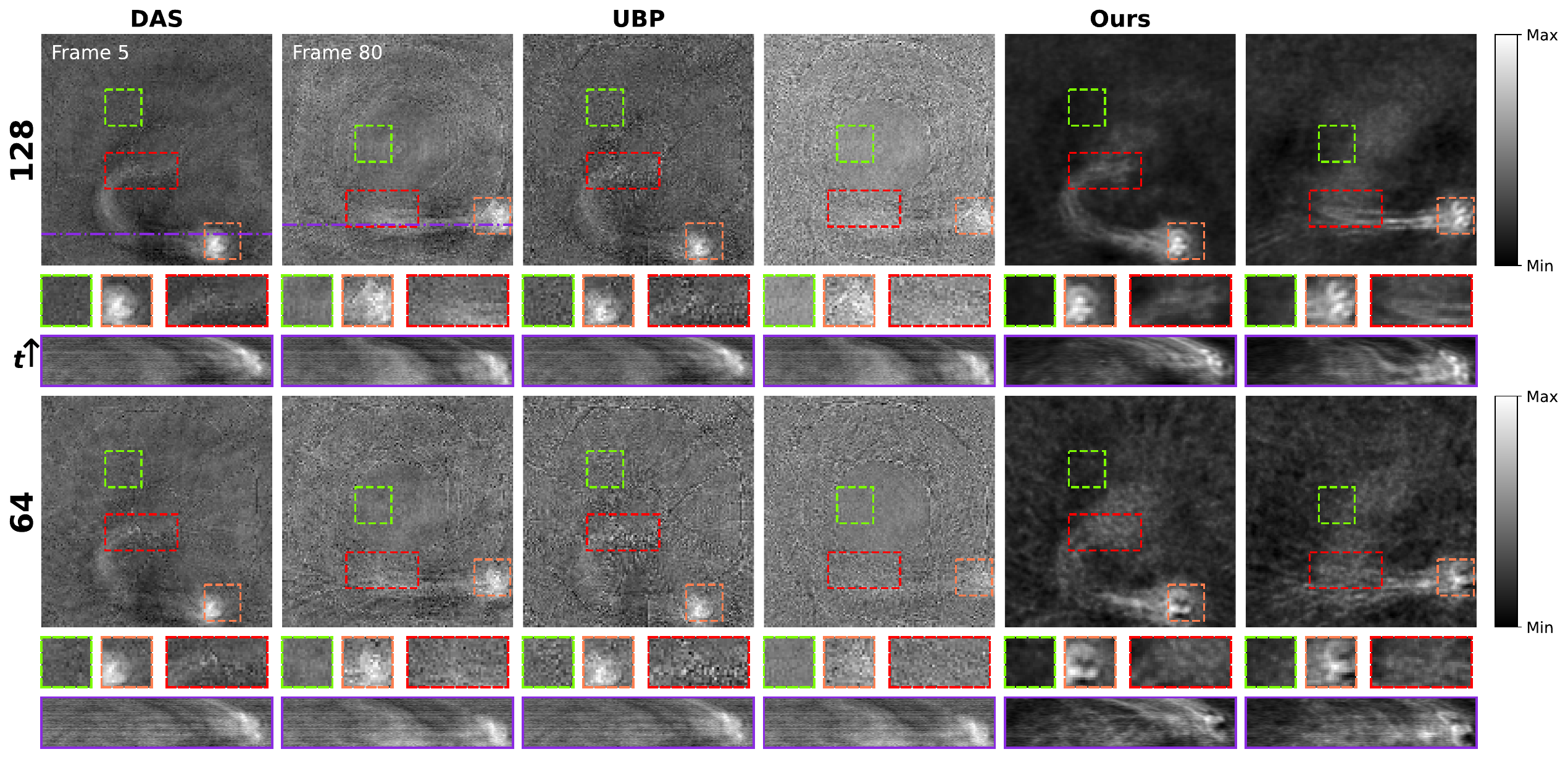}%
  \centering
  \caption{The reconstruction results on \textit{in vivo} fish data for DAS, UBP, and the proposed method (from left to right) are shown under sparse-view conditions with 128 and 64 sensors. The enlarged view of the fish head region is highlighted by orange boxes, and it is clearly observed that the proposed method is capable of reconstructing fine structural details. Error maps and average PSNR and SSIM metrics are displayed at the bottom.
  }
\end{figure*} 

\textit{2) In vivo Dynamic Finger Data:}
To validate the reconstruction capability of our method on real data, we first used a set of dynamically acquired PACT finger data. Since this data was collected using 128 sensors, it is inherently a sparse view, and no GT is available for comparison. Therefore, we can only perform a qualitative analysis of the reconstructed images. Fig. 4 shows the reconstruction results of different methods on this finger dataset. From the figure, it is evident that with 128 sensors, our method is able to reconstruct better vascular contours and shapes, especially for small vessels (e.g., the small vessels inside the red box), which are barely visible and difficult to distinguish in the DAS and UBP results. As the sparsity increases further, the reconstruction performance of DAS and UBP decreases, and the vascular shapes start to blur, with less sharp edges. To better illustrate this phenomenon, we selected a contour line passing through the vessels from the image within the red box and displayed it in Fig. 4. Despite the presence of significant noise in the raw photoacoustic data, and some noise in the image at 64 sensors, the vascular contours and shapes remain clearly visible in our reconstruction, demonstrating the effectiveness of our method on real data.

\textit{3) In vivo Dynamic fish Data:}
To further validate the reconstruction capability and generalizability of our method on real data, we performed reconstruction on \textit{in-vivo} dynamic PACT fish data. Similar to the previous dataset, this data was acquired with 128 sensors, making it a sparse-view dataset, and no GT is available for reference. In this case, we can only conduct a qualitative analysis of the reconstructed images. Fig. 5 shows the reconstruction results of different methods on the fish dataset. It is evident that, with 128 sensors, due to the higher laser frequency used in this dataset, the energy distribution is uneven, leading to noticeable bright flashes (see the supplementary video) and high noise levels in the DAS and UBP results. Additionally, we observe that the head of the fish in the reconstructed image appears blurred, making it difficult to discern specific details, such as the eyes (marked in the orange box). This phenomenon becomes even more pronounced with 64 sensors. In contrast, our method, utilizing low-rank and TV loss, effectively exploits the inter-frame information, greatly reducing noise and reconstructing better details. In the area highlighted by the orange box, we can clearly distinguish the fish's eyes, showcasing the advantage of our method in preserving fine details.

\subsection{Interpolation performance comparisons}
To validate the effectiveness of the proposed dynamic PACT representation in modeling temporal continuity, we designed and conducted an interpolation performance comparison experiment. In this experiment, we first optimized a temporally undersampled image sequence, then input the complete set of temporal coordinates to perform temporal interpolation using the INR-based method. In the specific experimental setup, we selected a 20-frame image sequence from the mouse abdominal PACT dataset. To reduce the impact of large inter-frame motion differences on interpolation accuracy, we trained the model using only one-quarter of the frames under a sparse setting with 64 sensors. As shown in Fig.~6(b), the interpolated images are highly consistent with the GT frames that were not used during training. This demonstrates the high fidelity of our method in temporal interpolation and its potential to further improve the frame rate of dynamic PACT.

\begin{figure}[h]
  \centering
  \includegraphics[width=3.5in]{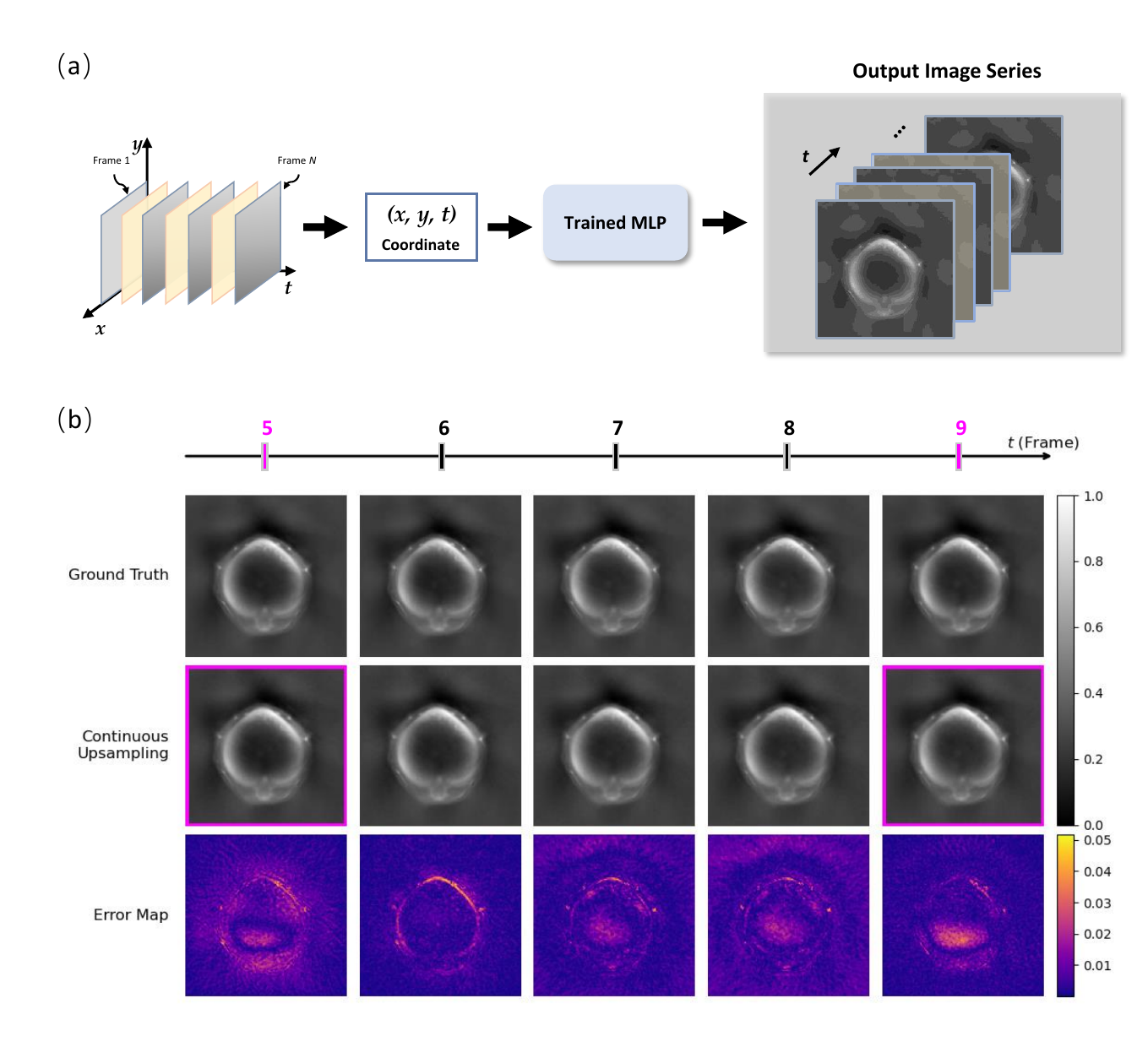}
\caption{(a) Temporal super-resolution pipeline for dynamic PACT reconstruction. Given a series of dense coordinate points, the optimized function utilizes an MLP to output interpolated frames, thereby enhancing the temporal resolution. (b) Upsampling between Frame 5 and Frame 9 of the \textit{in vivo} mouse abdominal dynamic PACT data, based on sparse views from 64 sensors. Three equally spaced time-point coordinates are generated and fed into the network to achieve 4x temporal super-resolution. The original GT frames serve as reference baselines. The frames used for training are highlighted with purple borders. The corresponding error maps are displayed below.
}
  \label{temporal_super_resolution}
\end{figure}

\section{Discussion}
For dynamic PACT, the number of sensor channels must satisfy the Nyquist sampling criterion to achieve high-quality image reconstruction. However, most existing dynamic PACT methods fail to leverage inter-frame temporal correlations, which can provide valuable information for improving reconstruction quality. In addition, the most widely used and effective sparse-view reconstruction methods are based on supervised deep learning, which require large amounts of training data and often suffer from poor generalization in real-world applications. Achieving high temporal resolution is crucial for practical PACT usage, yet the common approach relies on hardware upgrades, which substantially increase system costs. To address these challenges, we propose a novel unsupervised deep learning method based on INR for temporal super-resolution reconstruction under sparse-view conditions in dynamic PACT. The proposed method models the dynamic photoacoustic image sequence as a continuous spatiotemporal mapping function, effectively capturing both spatial and temporal information. We validated the proposed approach using simulated dynamic abdominal PACT data of mice, as well as \textit{in vivo} finger and zebrafish datasets under two different sparsity levels. Experimental results demonstrate the effectiveness and generalizability of our method in artifact suppression and motion fidelity. Remarkably, it can still recover structural details even under extremely sparse conditions. Compared to conventional dynamic PACT reconstruction methods such as DAS and UBP, our method consistently achieves superior performance, suggesting strong potential for high-temporal-resolution 2D photoacoustic imaging. The internal continuity of INR inherently provides implicit regularization, which underpins the superior performance of our method over baseline approaches. This is further confirmed by the 4× temporal super-resolution results shown in Fig.~6. Unlike existing super-resolution strategies, our INR-based approach does not require additional modeling or retraining; higher temporal resolution can be achieved simply by feeding denser time coordinates into the trained model. This significantly reduces computational burden and reconstruction time during deployment. Despite these advantages, real-time imaging remains an area for future improvement. Although our method is faster than other unsupervised approaches, it has not yet achieved real-time reconstruction. This limitation highlights the need for further research into computational optimization, potentially via algorithmic enhancements or hardware acceleration. Achieving real-time capability would greatly enhance the practical applicability of our method, particularly in clinical scenarios where immediate imaging feedback is critical.

\section{Conclusion}
In photoacoustic imaging, image reconstruction is a crucial step to obtain high-quality images. In this study, we propose an unsupervised deep learning method based on INR for temporal super-resolution dynamic photoacoustic image reconstruction under sparse views. The method learns an implicit continuous representation function that maps spatiotemporal coordinates to the corresponding image intensities, thereby modeling the spatiotemporal image sequence of the target. The proposed method does not rely on a training dataset and does not require any prior information for image reconstruction. We experimentally validated the method on simulated dynamic abdominal PACT data of mice, as well as real \textit{in vivo} finger and \textit{in vivo} fish data. The results show that even under sparse view conditions, the method can robustly generate high-quality dynamic photoacoustic image sequences. Furthermore, thanks to the internal continuity of the INR network, the method demonstrates a significant advantage in temporal super-resolution, enabling up-sampling of dynamic images at a temporal resolution higher than the physical acquisition rate. Therefore, we believe that the INR-based approach has the potential to further accelerate dynamic photoacoustic image acquisition in the future.
\bibliographystyle{unsrt}  
\bibliography{Main}        

\end{document}